# A two-dimensional space-time terahertz memory in bulk SrTiO$_3$

F. BLANCHARD[1,*], J. E. NKECK[1], L. GUIRAMAND[1], S. ZIBOD[2], K. DOLGALEVA[2], T. ARIKAWA[3], AND K. TANAKA[3]

[1] *Département de génie électrique, École de technologie supérieure (ÉTS), Montréal, Québec, Canada*
[2] *School of Electrical Engineering and Computer Science, Ottawa University, Ottawa, Canada*
[3] *Department of Physics, Kyoto University, Kyoto, Japan*
*\* francois.blanchard@etsmtl.ca*

Ferroelectric materials offer unprecedented ultrafast responses and are of great interest for the development of new polarizable media under the influence of an electromagnetic field. Recent research efforts have demonstrated the role of optical excitation and intense terahertz (THz) pulses in inducing a polar order and revealing a hidden phase transition in SrTiO$_3$ (STO), respectively. Here we show that the surface of STO crystals at room temperature act as ultrafast sensors that enable sub-picosecond switching through the Kerr effect and multi-ps recording of polar THz intensity with spatial resolution below the diffraction limit through dipole alignment relaxation. The contrast sensitivity and spatial resolution achieved by in the STO sensor are significantly superior to those of present-day near-field THz sensors based on the linear Pockels effect, and more importantly, its ability to remain polarized for several picoseconds opens the door to a new strategy for building an ultrafast space-time THz memory.

## Introduction

Ferroelectric memory has been commercially available for twenty years [1] and could play a key role in potential applications related to future 6G deployment [1,2]. Ferroelectric materials [3] possess a wide variety of responses to an external stimulus [4-8], such as electro-optic (EO) [4], piezoelectric [5], flexoelectric [6], magneto-optic [7], and piezomagnetic [8] responses, to name a few. Among ferroelectric materials, perovskite structures have emerged as promising and efficient low-cost energetic materials for various optoelectronic and photonic device applications [9]. To understand the response- and lifetimes of these materials, which range from nanoseconds to femtoseconds, ultrafast pump-and-probe spectroscopy techniques have proven to be an essential tool [10]. Their complex dynamics are evidenced by factors including the laser-induced ferroelectric structural phase transition observed by time-resolved X-ray diffraction [11], the laser-induced metastable ferroelectric state up to room temperature [12], the ferroelectric phase transition induced by an intense THz field revealed by time-resolved X-ray diffraction [13,14], and the ferroelectric phase transition induced by an intense THz field in STO measured by time-resolved optical spectroscopy [15].

The Kerr effect represents a common denominator for understanding the polarizability dynamics associated with electronic, vibrational and structural responses in ordered and disordered solids [16]. Recently, progress in the development of intense THz sources [17-19] has given rise to various revolutionary scientific demonstrations [20] such as the Kerr THz effect (TKE) [21-23]. Compared to the optical Kerr effect (or the AC Kerr effect), the TKE reveals the sign in a polarizable medium [23], much like the DC Kerr effect. Recently, TKE has been observed in bulk SrTiO$_3$ (STO) [15], a paraelectric material at room temperature, which is one of the most studied perovskite structures [12, 14, 15, 24-27]. Considering the estimated depth of penetration depth, on the order of micrometers (at THz frequencies in the bulk STO [15], limited by absorption [28,29]), a nonlinear interaction with this medium can only exist at its surface, where ferroelectricity can occur due to oxygen vacancy and/or impurity vacancies [30-33]. Therefore, similar to the case of the use of a thin EO sensor, where phase mismatch can be neglected [34-36], time-resolved near-field imaging by intense THz pumping could be envisioned to reveal the existence of highly localized information polarization at the STO surface. However, to the best of our knowledge, ferroelectric materials and/or the Kerr effect have never been reported as sensors for measuring, imaging or even recording structured THz waves below the diffraction limit.

Here, using intense THz pulses [18] coupled with a microscopic imaging capability [34], we reveal a new type of two-dimensional (2D) THz sensor based on the TKE and surface polarization of an STO bulk crystal. The ability to resolve THz images with sub-wavelength accuracy confirmed the surface origin of the detected signal. Unlike EO sensors based on the Pockels effect, whose sensitivity is linearly proportional to the electric field [37], the signal exhibits an orientational response due to a THz field-induced dipole alignment with a quadratic dependence and its subsequent slow decay. Using a specially designed metasurface for thin film material characterization [38], we confirmed the sign of induced polarization in the STO as well as its ability to accurately image THz waves with micron-level spatial resolution, primarily limited by the resolution of the optical probing beam [39]. Remarkably, our results indicate that the sign of the polarizability of the incident THz electric field can be confined within an area of a few tens of $\mu m^2$, thanks to our megapixel imaging method, and recorded for a duration longer than 10 ps. These results could lead to the development of a new strategy for ultrafast and ultra-dense two-dimensional (2D) THz memory, as well as the promotion of THz imaging applications [40], particularly for those with near-field capability, as a powerful technique to study ultrafast nonlinear light-matter interactions.

## Results

### THz Kerr effect in SrTiO$_3$

Fig. 1(a) illustrates the detection scheme used to observe the polar response in the STO at room temperature for (top) an optical probe and THz pump beam with polarization parallel to each other, and (bottom) perpendicular to each other. To realize this experiment (as detailed in the Material and methods section), pulses of 450 kV/cm at the focal point, and covering frequencies in the 0.1 to 3 THz range, were used. These THz waves are linearly polarized and emitted at a rate of 1 kHz by the tilted-pulse front pumping scheme (TPFP) in lithium niobate [18] (see also Fig. S1 in the Supplementary information). As can be seen in Fig. 1(b), the TKE signal from the STO crystal (red and blue curves) pumped by an intense THz field (black curve) changes polarity as a function of the orientation between the pump and probe polarization. In the same figure, we can also compare the sensitivity (in counts) between the Pockels effect in a 20 µm thick x-cut lithium niobate EO (from NanoLN) crystal (right side of the Y-axis) and the Kerr effect (left side of the Y-axis) in a 350 µm thick <100> cut bulk STO crystal (from MTI crystal). After taking into account the Fresnel losses for both crystals (see Materials and methods section), the detected signal in the STO is around 35 times smaller than that observed in a 20 µm thick LN crystal.

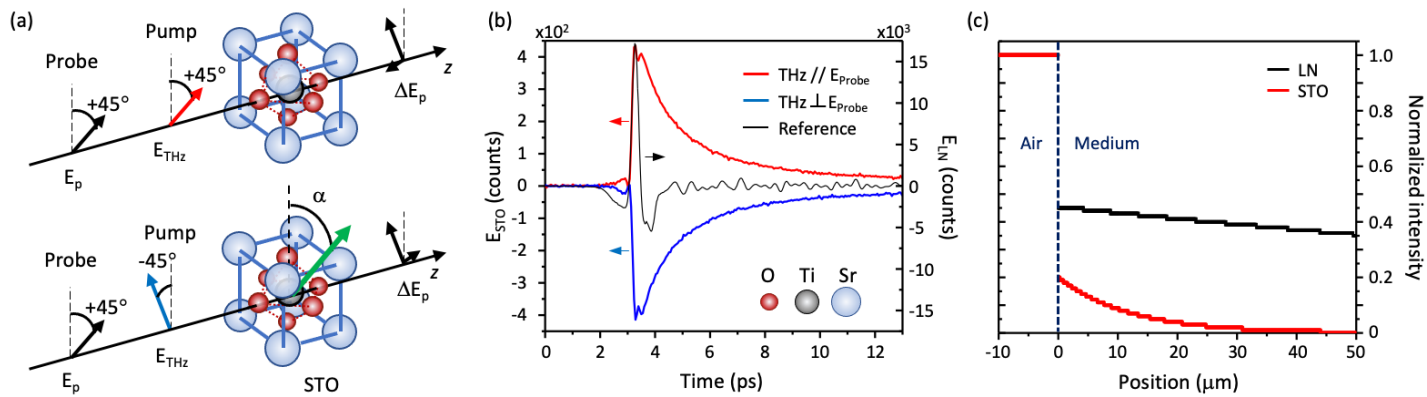

**Fig. 1. Experimental illustration of the THz pump - optical probe polarization.** Probe light condition (top) parallel and (bottom) perpendicular to the pump THz electric field. The green arrow shows the rotation $\alpha$ around the Z-axis. (b) The black curve is the THz signal obtained by EO sampling in lithium niobate ($E_{LN}$). The red and blue curves are the signals obtained by using the STO bulk material as a sensor ($E_{STO}$) for a probe light polarization parallel and perpendicular to the THz electric field, respectively. (c) Fresnel and absorption effects on THz intensity in LN and STO materials.

Indeed, because of the large absorption of THz waves inside the STO crystal [28,29], the interaction between the strong electric field and the STO must exist mainly in the first few microns, and not in the whole thickness range [15]. As shown in Figure 1(c), the THz intensity is drastically absorbed while traveling inside the STO crystal, with a loss of half its intensity after only 8 µm of propagation. In comparison in this same graph, the THz intensity in the LN crystal drops by less than 10% after 20 µm of propagation.

**Field and rotation dependency**

To clarify the STO response under an intense THz field, we investigated the detected signal dependency as a function of the crystal azimuthal angle. In this experiment, the THz beam propagates collinearly with a weak 800 nm probe beam initially linearly polarized at 45° to the THz polarization. While traveling collinearly in the STO crystal, the birefringence induced by the THz pump changes the polarization state of the probe pulse. Figure 2(a) shows the Kerr signal extracted at the peak position of the detected signal as a function of azimuthal angle $\alpha$ around the propagation Z-axis, as illustrated by the green arrow in Fig. 1(b). The observed sinusoidal behavior of the peak field as a function of the azimuthal dependence of the crystal agrees well with a phase retardation delay from the third-order polarization, as previously demonstrated for the TKE in gallium phosphide (GaP) [22].

To explore a regime of higher electric field values, a 4 mm diameter hemispherical silicon lens is inserted in direct contact with the sensor. This lens serves two purposes: to spatially confine the light to increase the local electric field at the detector surface and to improve the transmission factor of the THz wave between the air and the sensor (see the Supplementary Fig. S2). With this Si lens, the maximum incident field available at the sensor position reaches approximately 1.5 MV/cm. In order to evaluate the intensity dependency of the detected signal, two wire grid polarizers were used in the THz beam path to attenuate the field strength, by varying the angle of the first polarizer while keeping the second polarizer at a fixed angle. Figure 2(b) shows the detected waveforms for the STO sensor as a function of the incident electric field ranging from 330 kV/cm to 1.5 MV/cm. As expected in this figure, a linear dependence is observed for the LN sensor while a clear quadratic dependence is revealed for the STO sensor.

Finally, in Fig. 2(c), we plot the relaxation time of the dipole alignment for the high field condition (1.2 MV/cm) obtained using the Si lens, and with the lower field case (450 kV/cm), i.e., without the lens. Interestingly, in this figure, the behavior of the high field condition reveals a linear recovery time on a log-log plot, i.e., following a power-law dependence, whereas at low field, a faster and exponential recovery time is found. At first glance, this result would seem to indicate a change in behavior of the STO under the influence of an extremely high THz field, i.e., a possible transition from paraelectric to ferroelectric, much like in [15], but at room temperature. This important finding may also have its origin in our particular configuration, where the small silicon lens in contact with the STO crystal produces a different excitation from the free space focusing. The preceding notwithstanding, we considered that this notable result is beyond the scope of this work and should be addressed under comparable power dependence conditions. In keeping with the main objective of this paper, the next section focuses on the nature of the imaging obtained by the Kerr effect and the polarization state on the property of the THz images captured by the STO in comparison with that obtained with the Pockels effect commonly used for the EO detection of THz waves.

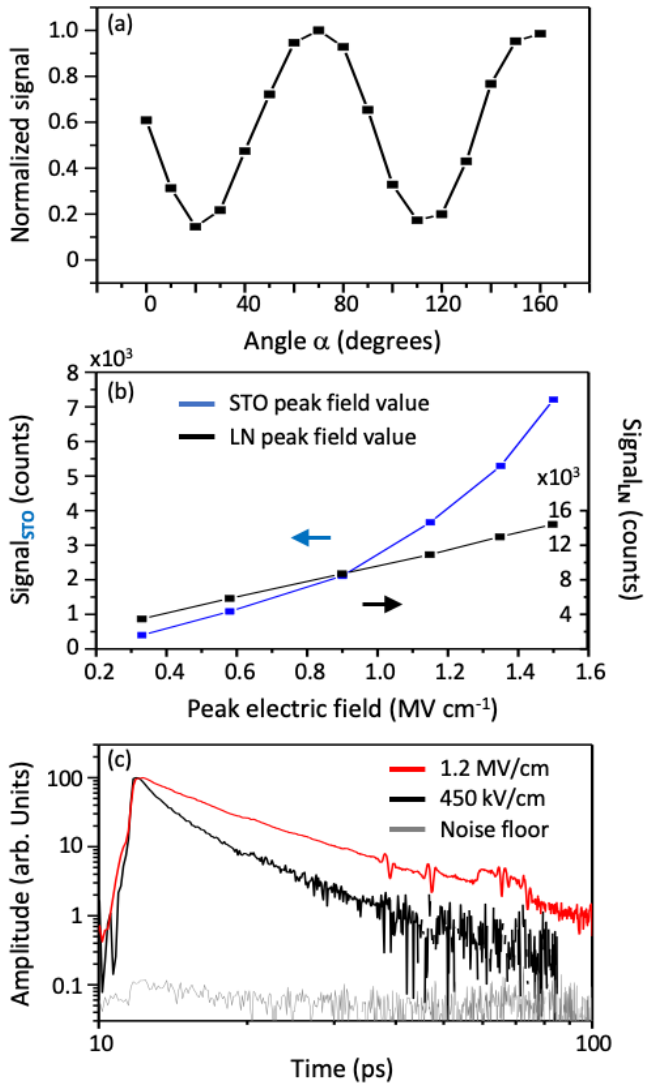

**Fig. 2. Characteristic responses of the STO crystal under intense THz pulse radiation.** (a) Azymuthal dependence around the z-axis. (b) Quadratic and linear dependence of the peak incident electric field for STO and LN sensors, respectively. (c) Time dependence of the dipole alignment relaxation on a log-log plot for an extremely high THz field using a 4 mm diameter Si lens (red curve) and without Si lens (black curve).

**THz microscopic imaging of a metasurface**

Recently, we introduced a method to characterize the optical properties of EO thin film materials using a THz near-field imaging technique [38]. The method is based on resolving the electric and/or magnetic field distributions in the near-field region of a specially designed split ring resonator (SRR) array for the THz frequency range, see an illustration of the SRRs in Fig. 3(a) and the experimental scheme in (b). From the response observed in the STO under the application of an intense THz field (as in Figs. 1(b) and 2(c)), we find that the electric field is unable to follow the fast oscillations after the passage of the strongest electric field. In other words, the fast oscillations following the main peak are averaged as a function of time because they are faster than the time needed for the polar orientation of the crystal to return to its initial condition. This phenomenon alters the space-time images of the SRRs on the STO following the passage of the intense THz wave, which should normally oscillate at its own resonance frequency [38].

In Fig. 3 (b), we present a snapshot of the simulated two-dimensional field lines (in yellow) obtained from the three-dimensional vector electric field for a single SRR excited by a vertically polarized THz wave (see also the Supplementary Fig. S3). In this figure, the cumulative maximum intensity distribution for the first picosecond of excitation is superimposed on this map to reproduce the spatial storage effect by STO polarization. It can be seen that the localization of the intensity clearly appears on the edges and on the corners of the SRR. With the help of the field lines, we note that the direction of the electric field on the four corners are opposite and perpendicular to each other, i.e., the electric field enters and leaves through the edges of the structure. To better visualize the field line orientation, + and - signs have been added in the areas where the field is aligned or perpendicular to the optical probe polarization, respectively.

Figures 3(d) and (e) show the THz near-field imaging maps at the best temporal position to reveal the sample shape for these two crystals, i.e., just after the peak of the THz signal, where the contrast between the reflected and transmitted waves is maximal. THz videos obtained with these two structures are shown in the supplementary material V2 and V3. The first distinction between the two THz images lies in the polar signature observed with the STO material. In the displayed time image of Fig. 3, for a sensor based on the Pockels effect, the observed image comes solely from the contrast between the THz field passing through the crystal and that reflected from the SRR wire. In the case of the STO sensor, the image comes from the same contrast, but with the addition of the localized intensity of the THz beam at the edge of the SRRs. The latter has a clear signature that follows the orientation of the field lines (in white and black) on the edges of the SRRs. This difference between the LN and STO sensors is particularly visible between the upper SRR image in Figs. 3(d) and 3(e).

Remarkably, even though the STO signal has a signal that is six times weaker than that of the 1 μm thick LN sensor, the resolving power and contrast remain higher. To emphasize this point, we show in Fig. 3(f) the line profile extracted between the two sensors at the upper SRR position, also corresponding to the position indicated by the red dashed line in Fig. 3(a). In Fig. 3(f), the spatial resolution is significantly improved (green curve) as compared to the 1 μm LN crystal (black curve), even though a bulk STO crystal is used. As reported previously, the spatial resolution degrades rapidly when a thick EO sensor is used [34,35]. Here, with the high-resolution image exceeding the diffraction limit captured by the STO, we can clearly confirm that the THz image is only due to a surface response of the STO, within a micrometer range.

The near-field image revealed by the STO sensor also confirms its ability to respond to different polarities with sub-wavelength accuracy. Basically, all SRRs show a polar intensity mapping under THz pulse excitation. For example, in the map in Fig. 3(e), the SRRs have a positive and negative polarity along their 50 μm long wire, as evidenced by the alternating black and white color on each SRR wire. Despite this very good spatial resolution performance, which is close to 1 μm for this THz microscope [39], we anticipate that the use of a thin STO crystal would further improve the THz Kerr imaging performance without compromising its sensitivity. An example of the strong induced birefringence inside the bulk STO is prenseted in Fig. S4.

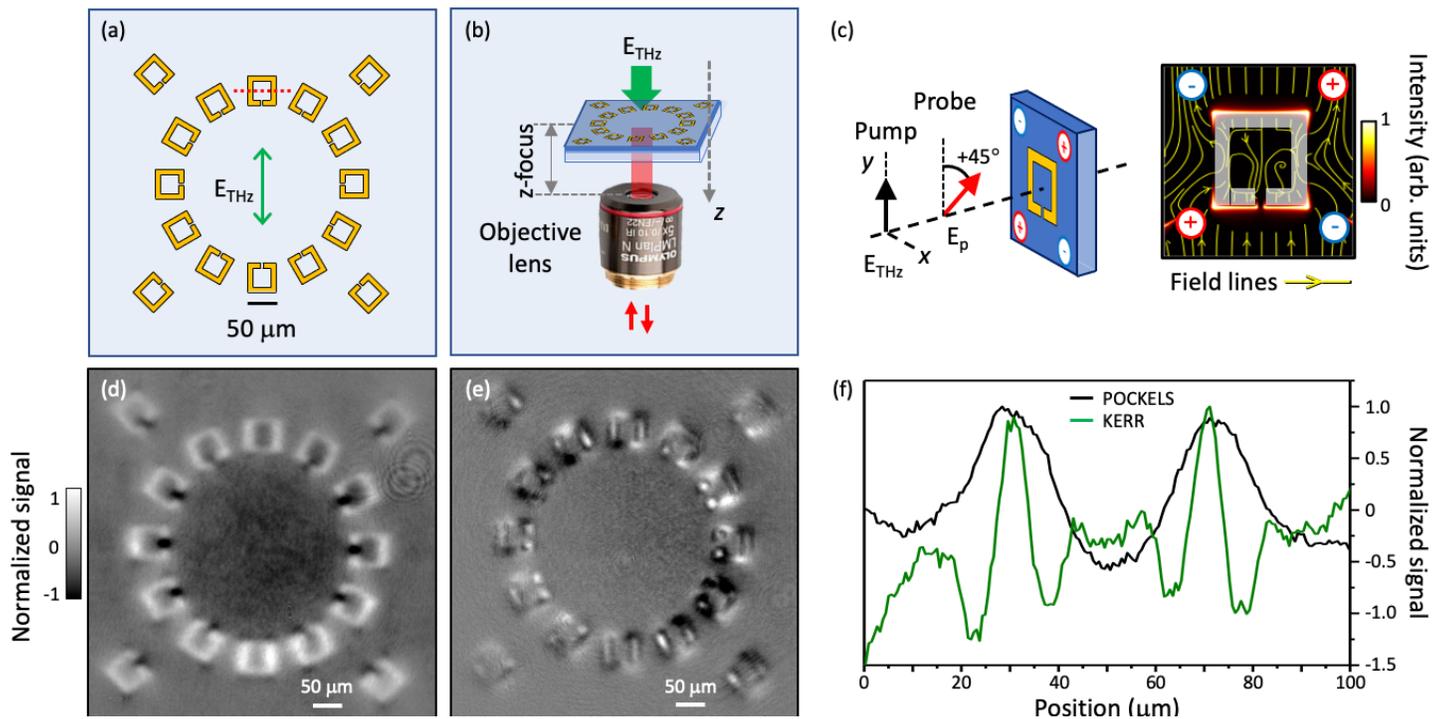

**Fig. 3. THz microscopic imaging of a metasurface.** (a) Illustration of the SRR pattern fabricated on the LN [38] and STO crystals. (b) Illumination scheme of the probe light on the EO sensors. (c) Illustration of a vertically polarized THz beam on a single SRR. The left side of the figure shows the simulated field lines and the expected polarity observed using a 45° probe light. The color scale represents the THz intensity. (d) EO imaging using a thin LN (see supplementary material video V1) crystal and (e) using a bulk STO crystal (see supplementary material video V2). (f) Extracted profiles for LN and STO imaging performance at the position of the red dotted line shown in (a).

**Spatiotemporal THz recording memory**

In the final step, we investigated the relationship between the slow decay response of the STO and near-field THz imaging performances. To differentiate the fast response (Kerr effect) from the slow decay (dipole alignment relaxation), we used a THz beam modulated by a vortex phase plate (VPP), as illustrated in Fig. 4(a). This phase plate is used to generate a 450 GHz vortex beam that exhibits a distinct doughnut-shaped structured intensity profile, as reported previously [36]. Unlike the SRR resonator, which oscillates in a highly localized manner, the vortex wave rotates around the same point without returning to the same location for a second time, i.e., the peak intensity moves continuously across the surface of the STO with a constant linear polarization. This distinction will allow us to evaluate the conservation of a structured THz beam through the alignment of the dipoles in the STO. Similarly with the results presented in Fig. 3, we compared the imaging performances between LN and STO.

Fig. 4(b) presents snapshots of the 450 GHz vortex beam passing through a 1 μm thick LN crystal. The yellow arrow indicates the direction of rotation of the vortex beam and the first 3 maps show the time range during which the vortex rotates around its central position. Indeed, after the passage of the vortex beam, no information other than water vapor appears in the detected image maps, as expected. On the other hand, after the first 3 maps shown in Fig. 4(c) using the STO, a stable doughnut-shape is "imprinted" on the surface of the STO, even after more than 10 ps.

In our imaging observations with the STO, it was found that a slow-moving wave is launched on the crystal surface when a sub-wavelength object or metal edge is in contact with the sensor and irradiated by the incident THz light, as in the case of metasurface imaging (see video V2 of metasurface by STO). Such waves travel at a very slow speed on the STO surface, in the order of 10 μm/ps, due to the very high real part of the refractive index of the $SrTiO_3$ in the THz frequency range [41]. In this case, by moving away from the object to be imaged, the spatial shape recorded in the STO slowly degrades. Interestingly, for the vortex beam, where the spatial shape of the THz beam is formed by a VPP before reaching the sensor, no surface waves traveling on the STO are observed. This information might be crucial since the local polarization in the STO medium remains intact in the case of structured THz light carrying subwavelength information.

**Discussion**

Similar to nonlinear optical microscopy (NLOM) [42], which uses optical nonlinearity to improve the contrast and sensitivity, the use of the nonlinear quadratic Kerr effect THz response is a promising alternative to achieve micrometer resolution imaging of low-contrast $mm^2$ size samples, such as biological cells. The idea of using a nonlinear material to enhance the contrast of THz images is not new [43], but having a sensor that inherently has this nonlinearity is more practical for microspectroscopic imaging.

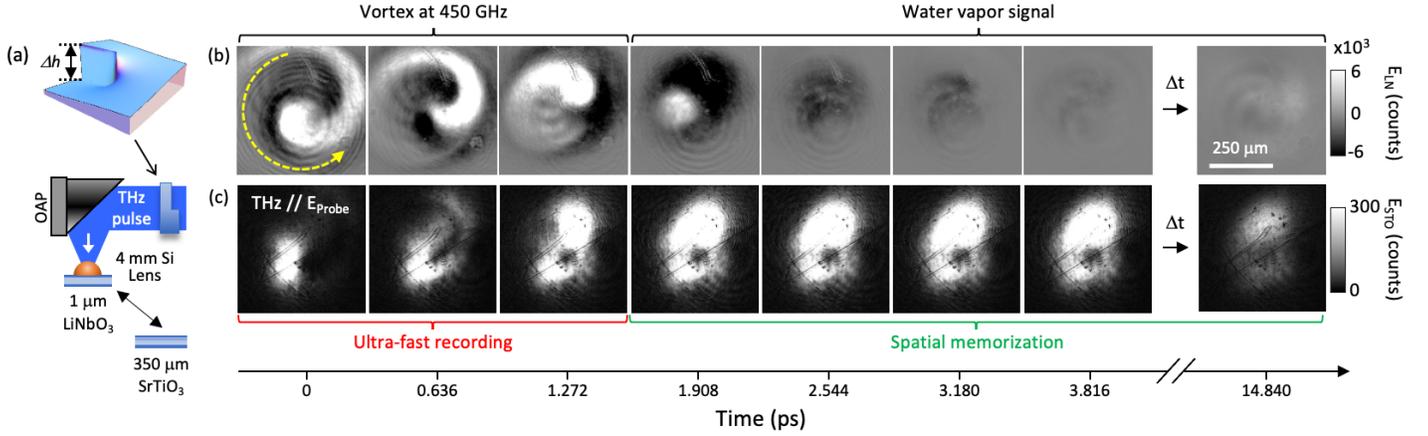

**Fig. 4. Spatiotemporal recording of THz vortex beam.** (a) Illustration of the experimental scheme. (b) THz vortex beam imaged using a 1 μm thick LN sensor (see supplementary material video V3) and (c) using a 350 μm thick bulk STO crystal (see supplementary material video V4). The selection of maps presented here is taken from the complete THz videos available in the supplementary material.

Furthermore, our results suggest that free-space recording of an upstream structured vortex beam does not suffer from THz diffraction at the surface of the STO that could blur the quality of the recorded image. Given the importance of using twisted light for information transport, and especially quantum cryptography [44,45], interest in recording waves with orbital angular momentum should continue to grow.

Finally, this work clearly demonstrated the writing of a megapixel in a few ps, followed by several ps of recording, which is orders of magnitude faster than the performance of typical RAM memories in common computers. With recent advances in the nonlinear control of material lattices by optical excitation [12], this type of space-time encryption should eventually allow ultrafast recording of THz waves, with information retention on much larger time scales, thus opening the door to ultrafast storage of massive amounts of data.

## Materials and methods

### Experimental details

Figure S1 schematically illustrates the experimental setup. We used a Ti:Sapphire mode-locked regenerative amplifier (Solstice, Spectra-Physics) that delivers 100 fs optical pulses (center wavelength, 780 nm) with a pulse energy of 3.2 mJ at a repetition rate of 1 kHz. Linearly polarized coherent THz pulses with a Gaussian beam profile were generated by optical rectification in a LiNbO$_3$ (LN) crystal using a tilted pulse front excitation scheme. For the Pockels effect THz pulse measurement in Fig. 1, we used a 20 μm thick x-cut LN mounted on a 1 mm thick z-cut LN substrate. The 2D Pockels effect EO imaging in Figs. 3 and 4 were performed from a 1 μm thick x-cut LN crystal mounted on a 500 mm glass substrate. The Kerr effect EO measurements were performed with a 350 μm thick STO <100> crystal. EO sampling was performed using a probe beam imaged by either a 5x or 10x lens on the EO crystal and relayed through a 16-bit PCO complementary metal oxide solid state camera (model PCO 5.5 edge) with a polarization analysis unit. EO sampling in the reflection geometry is achieved through a treatment of the top and bottom surfaces of the sensors coated with high reflection (HR) and anti-reflection (AR) films for the probe pulse at 780 nm, respectively. In the vortex beam experiments, we place an SPP in the collimated portion of the THz beam propagating in the z direction. The VPP is nominally designed to convert the Gaussian beam to a vortex beam (±ℏ) at 0.45 THz. The pitch height is 1.29 mm. The total step heights shown above are discretized into 16 small steps. In this experiment, to reduce the size of the vortex image and to obtain a higher intensity at the focal point, a 2 mm thick silicon hemispherical lens with a radius of 2 mm was used and placed directly in contact with the sensor. See the effect of Si lenses on microscopic images in Fig. S2.

### Sample fabrication

During the experiments with the SRR metasurface, gold SRR patterns were fabricated on the top surface (high-reflection coating material) of the THz detector crystal (LN and STO) using a standard photo-lithographic technique. Chromium (10 nm thick) was deposited under the 100 nm thick gold as an adhesion layer.

### FDTD simulations

We performed 3D simulations using Lumerical's finite difference time-domain (FDTD) software and extracted the $E_x$, $E_y$ and $E_z$ field components. A THz Gaussian beam with a pulse width of 0.35 ps and a center frequency of 1 THz was used to excite the SRRs, with its field polarized in the y direction. All simulation movies covered a time evolution of 20 ps, with a step resolution of 16.8 fs. A 500 mm thick glass substrate, a 1 mm thick lithium niobate crystal, and a 3 μm thick $SiO_2$ material for the high-reflectivity coating were included in the simulated materials, as indicated in the experimental conditions. In the region of SRR structures, the grid of the simulations was resolved to an accuracy of 0.1 mm. Figure S3 presents the three-dimensional vector field for 5 structures with a zoomed view on the vector field present at the corner of an SRR. To access the near-field information, a 2D monitor was placed 3 μm below the SRR position, i.e., at the interface between the HR coating and the sensor.

## Fresnel losses and Kerr effect in STO

To confirm the surface effect of the Kerr effect in STO, we theoretically calculated the transmitted THz electric field along the propagation direction, see Fig. 1(a). For that, we first took the Fresnel losses into account at the air-STO interface, and we considered the STO refractive index at 1 THz equal to 18 [1]. The transmission coefficient is given by: $t_{THz}^{STO} = \frac{2}{1+n_{STO}^{THz}} = \frac{2}{1+18} \simeq 0.105$. We also considered the STO absorption coefficient ($\alpha$), which is 420 cm$^{-1}$ at 1 THz [46]. Thus, the transmitted THz electric field along the propagation direction ($z$) in STO is given by: $E_{STO}^{THz}(z) = E_0^{THz} \cdot t_{THz}^{STO} \cdot e^{-\alpha z}$ with: $E_0^{THz}$ the incident THz electric field strength (1.2 MV/cm). We also compared this result with the transmitted THz electric field in LN along the propagation direction. The LN refractive index is 6.65 and the absorption coefficient is 25 cm$^{-1}$ at 1 THz [47]. From this calculation, we confirmed the low contribution of the electric field inside the STO for the THz Kerr. In comparison, the Pockels effect in the LN is not just a surface effect because it propagates with relatively little loss in the first 20 microns. Moreover, the THz electric field is 2.5 less transmitted and much more attenuated in the STO than in the LN, as shown in Fig. 1(c).

**Acknowledgments.** The authors acknowledge useful discussions with Mr. A. Doi, Professor K. A. Nelson and Professor A. M. Rappe.

**Funding.** F.B. gratefully acknowledges financial support from NSERC (2016-05020), the Canada Research Chair (CRC-2019-127) and the MRI project 2021-2022.

**Disclosures**. The authors declare no conflicts of interests.

**Data availability.** Data underlying the results presented in this paper are not publicly available at this time but may be obtained from the authors upon reasonable request.